\documentstyle[aps,preprint]{revtex}
\begin{document}

\title{
Numerical evidence for the spin-Peierls state in the frustrated
quantum antiferromagnet
}

\author{P. W. Leung and Ngar-wing Lam}
\address{Physics Department,
Hong Kong University of Science and Technology, Clear Water Bay,
Hong Kong
}
\date{\today}
\maketitle

\begin{abstract}
We study the spin-$1\over2$
Heisenberg antiferromagnet with an antiferromagnetic $J_3$
(third nearest neighbor) interaction on a square lattice.
We numerically diagonalize this
``$J_1$-$J_3$'' model on clusters up to 32-sites and search
for novel ground state properties as the frustration parameter
$J_3/J_1$ changes.  For ``larger'' $J_3/J_1$ we find
enhancement of incommensurate spin order, in agreement
with spin-wave, large-$N$ expansions, and other predictions.  But for
intermediate $J_3/J_1$, the low lying excitation energy spectrum
suggests that this incommensurate order is short-range.
In the same region, the first excited state has the symmetries
of the columnar dimer (spin-Peierls) state.
The columnar
dimer order parameter suggests the presence
of long-range columnar dimer order.
Hence, this spin-Peierls state is the best candidate for the
ground state of the $J_1$-$J_3$ model in an intermediate
$J_3/J_1$ region.
\end{abstract}
\pacs{75.10J, 75.40M}

The phase diagram of the frustrated spin-$\frac{1}{2}$
Heisenberg antiferromagnet has received much interest in recent years.
On a square lattice, frustration can be introduced by
further-than-nearest neighbor antiferromagnetic couplings.
Short-range interactions up to a distance of
two lattice constants have been studied.  This is the
``$J_1$-$J_2$-$J_3$'' model, which is described by the Hamiltonian
\begin{equation}
{\cal H}=J_1\sum_{nn}{\bf S}_i\cdot{\bf S}_j +
J_2\sum_{2nn}{\bf S}_i\cdot{\bf S}_j +
J_3\sum_{3nn}{\bf S}_i\cdot{\bf S}_j,
\end{equation}
where the sums run over all first, second, and third nearest neighbors,
and all $J_i>0$.
The classical phase diagram of this model is well-known to have
transition lines between N\'eel, collinear, and spiral states
\cite{gsh89}.
The critical line separating the N\'eel and spiral states,
$J_1-2J_2-4J_3=0$, is called the classical critical line (CCL).
The quantum phase diagram is less clear.
When the frustration is small (at small $J_2$ and $J_3$), the
model possesses N\'eel order.
Various analytical studies
including linear spin-wave\cite{spinwave,mdjr90}
and mean-field theories \cite{meanfield}
have shown that the ground state possesses
collinear and spiral (incommensurate) spin order at large $J_2$ and $J_3$
respectively.
The ground state at intermediate $J_2$ and $J_3$, particularly along the
quantum analog of the CCL,
is still controversial.
While some theories \cite{spinwave,mdjr90,il88}
predict that frustration and quantum fluctuation
destroy the N\'eel order to form a state without spin order,
others \cite{meanfield,f93}
predict that quantum fluctuation
can stabilize the N\'eel state along this critical line.
Nevertheless, it seems likely that in between the N\'eel
and spiral phases, there exists an intermediate state without spin order.
Large-$N$ expansions of the unfrustrated antiferromagnet
\cite{meanfield_unfrustrated} predict
that this intermediate state
is spontaneously dimerized.  This is supported by
series expansion on the $J_1$-$J_2$-$J_3$ model \cite{gsh89},
although the former (Ref.\cite{meanfield_unfrustrated})
did not treat the frustration due to $J_2$
and $J_3$ explicitly.
On the other hand, spin-wave theory \cite{spinwave,mdjr90,f93}
predicts that this intermediate state
is a spin-liquid.

The search for a spin-liquid state in low-dimensional quantum antiferromagnets
has long been a fascinating problem.
Such a state is most likely to be found in frustrated systems with large
quantum fluctuations.  Therefore, the region along the
quantum analog of the CCL in the
$J_1$-$J_2$-$J_3$ model is a good place to search
for a spin-liquid state, although
other kinds of long-range order (such as spin-Peierls order)
have been proposed in the same region.
Since the location of this critical line
is unknown, it is tedious to work with two adjustable
parameters $J_2$ and $J_3$.
Most numerical diagonalization studies of the frustrated Heisenberg
model have been for $J_3=0$ \cite{pgbd91,j1j2}.
A recent study \cite{f93} has suggested that the end of the critical
line on the $J_2$ axis is a Lifshitz point and thus not representative
of the whole critical line.
The purpose of this letter
is to study this model on the  $J_3$ axis.
In the following, we will take $J_1=1$ and $J_2=0$.

Using the Lanczos algorithm, we are able to diagonalize this $J_1$-$J_3$
model on a 32-site square cluster.
Most results in this letter are obtained from the 16-site
(4$\times$4) and 32-site square lattices.
It is obvious that the 16-site lattice is too small to include
the $J_3$ interaction because each site has only two (instead of four)
distinct third
nearest neighbors.
Hence we will use results obtained from the 16-site
system for comparison only.  Finite-size scaling of
the 16- and 32-site results will not be reliable,
except perhaps at small $J_3$.

It is well known that for $J_3=0$, the ground state of the
$J_1$-$J_3$ model
exhibits long-range N\'eel order \cite{m91}.
Fig.~\ref{fig_md} shows the finite-size plot of the staggered magnetization
$m^\dagger$, defined as
$m^{\dagger2}=S(\pi,\pi)/N$ (see Eq.~\ref{sq}).
The system sizes are $N=16$, 24 \cite{N24}, and 32.
The $1/\sqrt{N}$ dependence
is taken from spin-wave theory for the unfrustrated case \cite{nz89}.
We can see that at $J_3=0.3$, $m^\dagger$ extrapolates to a finite value
as $N\rightarrow\infty$.  But the linear extrapolation fails at $J_3=0.35$.
Hence we conclude that the N\'eel order persists at least up to $J_3=0.3$
in the thermodynamic limit.
To study the spin order as $J_3$ increases further, we calculate
the static structure factor,
\begin{equation}
S({\bf q})=\frac{1}{N}\sum_{kl}e^{i{\bf q}\cdot({\bf R}_k-{\bf R}_l)}
\langle{\bf S}_k\cdot{\bf S}_l\rangle. \label{sq}
\end{equation}
Fig.~\ref{fig_sq} shows $S({\bf q})$
for the 32-site square lattice at different $J_3$.
It clearly shows that as $J_3$ increases from zero, the peak
shifts from $(\pi,\pi)$ to $(\frac{3\pi}{4},\frac{3\pi}{4})$ at $J_3\sim0.5$,
and then to $(\frac{\pi}{2},\frac{\pi}{2})$ at $J_3\sim0.7$.
This shows that the N\'eel order vanishes as $J_3$ increases
and another spin order
develops which has ordering vector along the $(1,1)$ direction.
If the system possesses incommensurate spin order,
the peak in $S({\bf q})$ should shift continuously from
$(\pi,\pi)$ to $(\frac{\pi}{2},\frac{\pi}{2})$ as $J_3$ increases.
Due to the discrete nature of the cluster,
such a continuous shift along the $(1,1)$ direction is not possible.
Nevertheless, Fig.~\ref{fig_sq} does indicate that incommensurate
spin order develops as $J_3$ increases.
We also calculate the dynamic structure factor \cite{pgbd91},
$S({\bf q},\omega)$.  Sharp low energy peaks are found at momenta
along the $(1,1)$ direction.
As $J_3$ increases, the lowest energy peak changes from $(\pi,\pi)$
to $(\frac{3\pi}{4},\frac{3\pi}{4})$ at $J_3\sim0.5$,
and then to $(\frac{\pi}{2},\frac{\pi}{2})$ at $J_3\sim0.7$,
indicating that the N\'eel order vanishes and the system
develops another spin order which has ordering vector
along the $(1,1)$ direction.

To study whether this incommensurate spin order is long-range,
we calculate the twist correlation function \cite{ccl90},
\begin{equation}
\chi_t=\biggl\langle\biggl |\frac{1}{N}\sum_{\bf r}
{\bf S}_{\bf r}\times({\bf S}_{{\bf r}+{\bf x}}+{\bf S}_{{\bf r}+{\bf y}})
\biggr|^2\biggr\rangle,
\end{equation}
where {\bf x} and {\bf y} are unit vectors.
We expect $\chi_t$ to be independent of $N$ for large enough $N$
if the system possesses long-range incommensurate spin order.
Fig.~\ref{fig_chitd}(a) shows $\chi_t$ in the 16- and 32-site systems
at different $J_3$.  In both systems, $\chi_t$ is enhanced
at $J_3$ larger than about 0.4.  This enhancement suggests the existence
of incommensurate spin order at large $J_3$, which is consistent with
the findings from the static and dynamic structure factors
discussed above.  However, only spin-wave excitations will show up
as peaks in $S({\bf q},\omega)$.  Therefore, it
does not exclude the existence of singlet excitations, especially
at intermediate $J_3$.
In particular it is difficult to judge from
Fig.~\ref{fig_chitd}(a) whether the incommensurate order is long-range
at intermediate $J_3$.

If a system possesses a broken symmetry
in the thermodynamic limit, the ground state of the finite
system will still be fully symmetric.  In this case the
ground state expectation of the appropriate
order parameter will have long-range correlations,
and there will exist low lying excited states
with the appropriate symmetries whose energy gaps vanish
in the thermodynamic limit \cite{hl88}.
Consequently, we can use the low lying energy levels of a finite
system to study the possible existence of long-range order.
Fig.~\ref{fig_energy} shows the energies of a few low lying
eigenstates in the 32-site system.  For $J_3$ smaller than about
$0.4$, the first excited state is a triplet with momentum $(\pi,\pi)$,
consistent with the existence of N\'eel order for small $J_3$.
We denote this state as $E_{T1}$.
For $J_3$ larger than about $0.85$, the first excited state
is a triplet with momentum $(\frac{\pi}{2},\frac{\pi}{2})$.
We denote this state as $E_{T2}$.  $E_{T1}$ and $E_{T2}$ are
the two spin-wave excitations which show up as low energy
peaks in $S({\bf q},\omega)$.
However, at intermediate $J_3$,
states with momentum $(\frac{3\pi}{4},\frac{3\pi}{4})$
are never the first excited state.
We denote the first excited
state in this region as $E_{S}$.
It is a  two-fold degenerate singlet pair, one
with momentum $(0,\pi)$ and the other with
$(\pi,0)$.  Both are odd under reflection along the direction
orthogonal to their momenta.
The symmetries of the $E_{S}$ state resemble the columnar dimer
state \cite{rk88}.  In this state, nearest neighbor spin pairs
form singlets (dimers), and these dimers freeze into a columnar order.
It is four-fold degenerate, and can form four states
with distinct symmetries: two with zero momentum, of which one is
fully symmetric while the other is odd under rotation;
and two with momenta $(0,\pi)$ and $(\pi,0)$.
The last two have the same symmetries as the degenerate $E_{S}$ state.

Since the finite system always has a first excited state no matter
whether the ground state possesses true long-range order, the
results of the above study
of the low lying states alone are not sufficient to show the
existence of dimer order.  The next evidence comes from
the order parameter for the columnar dimer state \cite{s89},
\begin{equation}
\theta^{dim}_{\bf r}=(-1)^{r_x}\,{\bf S}_{\bf r}\cdot{\bf S}_{{\bf r}+{\bf x}}
+i\,(-1)^{r_y}\,{\bf S}_{\bf r}\cdot{\bf S}_{{\bf r}+{\bf y}}.
\end{equation}
In finite-size calculations, we examine the correlation function
\begin{equation}
\chi_{dim}=\biggl\langle\biggl|\frac{1}{N}\sum_{\bf r}
\theta^{dim}_{\bf r}\biggr|^2\biggr\rangle.
\end{equation}
If the ground state has long-range columnar dimer order,
$\chi_{dim}\sim O(1)$.
Fig.~\ref{fig_chitd}(b) shows $\chi_{dim}$ at various $J_3$
in the 16- and 32-site systems.  Both systems have a peak in $\chi_{dim}$,
indicating that columnar dimer order is enhanced in the
corresponding  region
of $J_3$.  In the 32-site system, the peak is at $J_3\sim 0.7$,
which corresponds to the minimum energy gap between $E_0$ and $E_S$
in Fig.~\ref{fig_energy}.  The different peak positions
in Fig.~\ref{fig_chitd}(b)
may be due to finite-size effect of the 16-site lattice as discussed
above.  This effect also prevents us from doing
a reliable finite-size
scaling study of the peak values of $\chi_{dim}$.

The dimer correlations can be demonstrated clearly by calculating the
dimer-dimer correlation function \cite{le93} defined as
\begin{equation}
C_{(i,j)(k,l)} =\langle({\bf S}_i\cdot{\bf S}_j)({\bf S}_k\cdot{\bf S}_l)
\rangle-\langle{\bf S}_i\cdot{\bf S}_j\rangle^2,
\end{equation}
where the bracket $(m,n)$ denotes nearest neighbor sites.
A dimer liquid state
will display short-range structure in $C_{(i,j)(k,l)}$ but
decrease to zero at large dimer separations.
On the other hand, a dimer solid, or spin-Peierls state,
will continue to show periodic oscillations reflecting the underlying
long-range order.  $C_{(i,j)(k,l)}$
for all inequivalent dimer pairs of the 32-site lattice evaluated at $J_3=0.7$
are tabulated in Table~\ref{tab_dimer}.
Fig.~\ref{fig_dimer} is a pictorial representation of
$C_{(i,j)(k,l)}$.
The reference bond $(i,j)$ is represented by a double line.
For all other bonds $(k,l)$, the magnitude of $C_{(i,j)(k,l)}$
is represented by the thickness of the line joining sites $k$ and $l$.
Solid lines represent positive correlation, and broken lines
represent negative or anti-correlation.  It is clear that
nearest neighbor spin pairs tend to form dimers, and the
dimers are arranged in a columnar fashion.  The dimer-dimer
correlations do not decrease appreciably in the largest dimer
separation allowed in our system size.

To conclude, our numerical results show that N\'eel order
in the $J_1$-$J_3$ model is stable up to $J_3>0.3$, as compared
to $J_3=0.25$ in the classical case.  This could be the result of the CCL being
moved to much larger $J_2$ and $J_3$ values by quantum fluctuation \cite{f93},
when the N\'eel state is stabilized (order from disorder) and
the spiral state is destabilized along the critical line.
Our results further show that the model
is likely to have a spin-Peierls state between the N\'eel state
(at small $J_3$) and the incommensurate state (at large $J_3$),
in agreement with theoretical predictions
\cite{gsh89,meanfield_unfrustrated,sb90}.
In particular, Ref.\cite{meanfield_unfrustrated} predicted that
the dimerized patterns depend on the spin $S$, and our results
agree with it for $S=1/2$.  We would like to remark that
Ref.\cite{meanfield_unfrustrated} did not treat the frustration
explicitly, and it is not trivial that it gives the right prediction
along the quantum analog of the CCL.  But when the same analysis is extended
to include frustrations due to $J_2$ and $J_3$, similar dimerization
patterns are found \cite{tk}.
At large $J_3$, we believe, as our
results show, that the model possesses incommensurate spin order.
But our finite cluster has no wave vector from
$(\frac{3\pi}{4},\frac{3\pi}{4})$ to $(\frac{\pi}{2},\frac{\pi}{2})$
along the $(1,1)$ direction.  Hence we are not able to locate
the transition point to the incommensurate state nor to study
the order of the transition.

\bigskip
We thank T. K. Ng and R. J. Gooding for very useful discussions.
NWL acknowledges support from T.K. Ng and K. Y. Szeto through
the Hong Kong Research Grant Council (RGC) contract
number UST123/92E.  Part of the numerical diagonalizations
were performed on an HP workstation cluster at the Center for
Computing and Communication Services (CCST) of the Hong Kong University
of Science and Technology.  Technical support provided by the CCST
staff is gratefully acknowledged.


%
%
\begin{figure}
\caption{Finite-size plot of the
staggered magnetization $m^\dagger$ at different
$J_3$.  The straight lines are the best fit to the data.
The dotted lines are straight lines joining the data points.}
\label{fig_md}
\end{figure}

\begin{figure}
\caption{The static structure factor for the 32-site lattice
at different values of $J_3$.}
\label{fig_sq}
\end{figure}

\begin{figure}
\caption{(a) $\chi_t$ and (b) $\chi_{dim}$
at different $J_3$ in the 16-site ($\bullet$)
and 32-site ($+$) systems.}
\label{fig_chitd}
\end{figure}

\begin{figure}
\caption{Lowest energy states in different momentum sectors at
different $J_3$ in the 32-site system.  $E_0$ is the ground state.
$E_{T1}$ and $E_{T2}$
are spin triplets while $E_S$ is a spin singlet.  For the purpose
of clarity, lowest energy states in other momentum sectors
and the second lowest energy
state with momentum ${\bf q}=(0,0)$ are not shown.
These states have higher energy than the
second lowest energy states shown in the figure.}
\label{fig_energy}
\end{figure}

\begin{figure}
\caption{Dimer-dimer correlation function $C_{(19,23)(k,l)}$ of
the 32-site system at $J_3=0.7$.  The reference bond $(19,23)$
is represented by a double line.  The magnitude of $C_{(19,23)(k,l)}$
is proportional to the thickness of the line joining the pair
of sites $(k,l)$.  The solid line means $C_{(19,23)(k,l)}$
is positive, and the broken line means $C_{(19,23)(k,l)}$ is negative.}
\label{fig_dimer}
\end{figure}

%
%
\begin{table}
\setdec 00.00000
\caption{Dimer-dimer correlation functions for all inequivalent
dimer pairs in the 32-site system at $J_3=0.7$.
The reference pair is $(19,23)$.
See Fig.~5 for the
numbering of the sites.}
\label{tab_dimer}
\begin{tabular}{r@{}lcr@{}lc}
$(k,$&$l)$&$C_{(19,23)(k,l)}$&$(k,$&$l)$&$C_{(19,23)(k,l)}$\\
\tableline
(1,&5)       &\dec  0.055075     &(18,&22)        &\dec $-$0.054987 \\
(1,&29)       &\dec  0.003783    &(19,&22)        &\dec $-$0.019733 \\
(2,&29)       &\dec  0.051208    &(21,&25)        &\dec  0.003865 \\
(5,&10)       &\dec $-$0.051912  &(21,&26)       &\dec $-$0.045224  \\
(10,&14)       &\dec  0.067346   &(22,&26)        &\dec  0.009617 \\
(13,&18)        &\dec  0.051260  &(22,&27)        &\dec  0.104930 \\
(14,&19)       &\dec $-$0.071722 &(25,&29)       &\dec $-$0.042596 \\
(17,&21)       &\dec  0.050038   &(26,&29)       &\dec  0.003208 \\
(18,&21)        &\dec  0.002528  &(26,&30)       &\dec  0.067526 \\
\end{tabular}
\end{table}

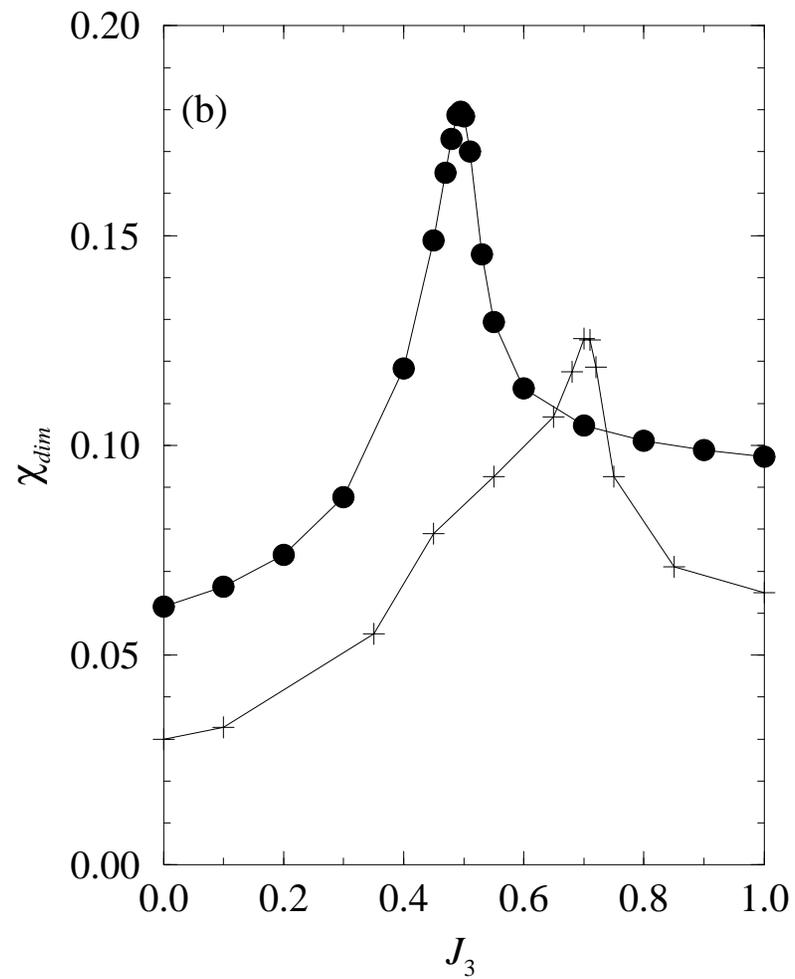
\end{document}